%% file: paper.tex
\documentstyle{mn}

\newif\ifAMStwofonts
\AMStwofontstrue

\newcommand{\gcc}{$g~cm^{-3}\ $}

\newcommand{\msun}{$M_{\odot}\ $}
\newcommand{\msune}{M_{\odot}\ }
\newcommand{\greq}{$\stackrel{>}{ _{\sim}}$}

\newcommand{\lteqe}{\stackrel{<}{ _{\sim}}}

\ifoldfss
  \ifCUPmtlplainloaded \else
    \NewTextAlphabet{textbfit} {cmbxti10} {}
    \NewTextAlphabet{textbfss} {cmssbx10} {}
    \NewMathAlphabet{mathbfit} {cmbxti10} {} 
    \NewMathAlphabet{mathbfss} {cmssbx10} {} 
  \fi
  \ifAMStwofonts
    \ifCUPmtlplainloaded \else
      \NewSymbolFont{upmath} {eurm10}
      \NewSymbolFont{AMSa} {msam10}
      \NewMathSymbol{\upi}     {0}{upmath}{19}
      \NewMathSymbol{\umu}     {0}{upmath}{16}
      \NewMathSymbol{\upartial}{0}{upmath}{40}
      \NewMathSymbol{\leqslant}{3}{AMSa}{36}
      \NewMathSymbol{\geqslant}{3}{AMSa}{3E}

       \let\le=\leqslant
       
    \fi
  \fi
\fi 

\ifnfssone
  \newmathalphabet{\mathit}
  \addtoversion{normal}{\mathit}{cmr}{m}{it}
  \addtoversion{bold}{\mathit}{cmr}{bx}{it}
  \newmathalphabet{\mathbfit} 
  \addtoversion{normal}{\mathbfit}{cmr}{bx}{it}
  \addtoversion{bold}{\mathbfit}{cmr}{bx}{it}
  \newmathalphabet{\mathbfss} 
  \addtoversion{normal}{\mathbfss}{cmss}{bx}{n}
  \addtoversion{bold}{\mathbfss}{cmss}{bx}{n}
  \ifAMStwofonts
    \ifCUPmtlplainloaded \else
      \UseAMStwoboldmath
      \makeatletter
      \new@mathgroup\upmath@group
      \define@mathgroup\mv@normal\upmath@group{eur}{m}{n}
      \define@mathgroup\mv@bold\upmath@group{eur}{b}{n}
      \edef\UPM{\hexnumber\upmath@group}
      \new@mathgroup\amsa@group
      \define@mathgroup\mv@normal\amsa@group{msa}{m}{n}
      \define@mathgroup\mv@bold\amsa@group{msa}{m}{n}
      \edef\AMSa{\hexnumber\amsa@group}
      \makeatother
      \mathchardef\upi="0\UPM19
      \mathchardef\umu="0\UPM16
      \mathchardef\upartial="0\UPM40
      \mathchardef\leqslant="3\AMSa36
      \mathchardef\geqslant="3\AMSa3E

       \let\le=\leqslant

    \fi
  \fi
\fi 

\ifnfsstwo
  \DeclareMathAlphabet{\mathbfit}{OT1}{cmr}{bx}{it}
  \SetMathAlphabet\mathbfit{bold}{OT1}{cmr}{bx}{it}
  \DeclareMathAlphabet{\mathbfss}{OT1}{cmss}{bx}{n}
  \SetMathAlphabet\mathbfss{bold}{OT1}{cmss}{bx}{n}
  \ifAMStwofonts
    \ifCUPmtlplainloaded \else
      \DeclareSymbolFont{UPM}{U}{eur}{m}{n}
      \SetSymbolFont{UPM}{bold}{U}{eur}{b}{n}
      \DeclareSymbolFont{AMSa}{U}{msa}{m}{n}
      \DeclareMathSymbol{\upi}{0}{UPM}{"19}
      \DeclareMathSymbol{\umu}{0}{UPM}{"16}
      \DeclareMathSymbol{\upartial}{0}{UPM}{"40}
      \DeclareMathSymbol{\leqslant}{3}{AMSa}{"36}
      \DeclareMathSymbol{\geqslant}{3}{AMSa}{"3E}

       \let\le=\leqslant

    \fi
  \fi
\fi 

\ifCUPmtlplainloaded \else
  \ifAMStwofonts \else 
    \def\upi{\pi}
    \def\umu{\mu}
    \def\upartial{\partial}
  \fi
\fi

\title[A General Relativistic Calculation]
{A General Relativistic Calculation of Boundary Layer and Disk Luminosity 
for Accreting Non--magnetic Neutron Stars in Rapid Rotation}
\author[A.V. Thampan and B. Datta]
       {Arun V. Thampan$^1$ \thanks{e--mail: arun@iiap.ernet.in}, 
       and Bhaskar Datta$^{1,2}\thanks{e-mail: datta@iiap.ernet.in}$\\
\parbox[t]{10cm}{$^{1}~$ Indian Institute of Astrophysics, Bangalore 560 034,
India}\\
\parbox[t]{10cm}{$^{2}~$ Raman Research Institute, Bangalore 560 080,
India.}}

\date{ }

\pagerange{\pageref{firstpage}--\pageref{lastpage}}
\pubyear{1997}

\begin{document}

\maketitle

\input psbox.tex

\label{firstpage}

\begin{abstract}
We calculate the disk and boundary layer luminosities for accreting 
rapidly rotating neutron stars with low magnetic fields in a fully 
general relativistic manner.  Rotation increases the disk luminosity 
and decreases the boundary layer luminosity. A rapid rotation of the neutron
star substantially 
modifies these quantities as compared to the static limit.  For a neutron star 
rotating close to  the centrifugal mass shed limit, the total luminosity has 
contribution only from the extended disk. For such maximal rotation rates, 
we find that much before the maximum stable gravitational mass configuration
is reached, there exists a 
limiting central density, for which  particles in the innermost stable orbit 
will be more tightly bound than those at the surface of the neutron star. 
We also calculate the angular velocity profiles of particles in Keplerian 
orbits around the rapidly rotating neutron star. The results are illustrated
for a representative set of equation of state models of neutron star matter.
\end{abstract}











\section{Introduction}

For accreting neutron stars in old binary systems, also known as low--mass
X--ray binaries (LMXBs), a narrow boundary layer girdling the neutron star 
will form next to the neutron star surface. The importance of the boundary 
layer derives from the possibility that this could be the site for the emission 
of a variable isothermal blackbody radiation component observed in the spectra 
of LMXBs characterized by very high X--ray luminosity (Mitsuda et al 1984). 
For weak magnetic field neutron stars,  the boundary layer is expected to  be 
substantially more X--ray luminous than the entire extended accretion disk 
on general theoretical grounds (Sunyaev \& Shakura 1986; King 1995). 
An important feature of disk accretion onto a weakly magnetized neutron star is 
that the neutron star will get spun up to its equilibrium period ($\sim$ 
milliseconds), over a timescale of hundreds of millions of years (see
Bhattacharya \& van den Heuvel 1991 and  references therein).  A rapid spin 
of the neutron star will enhance its equatorial radius and also relocate the 
inner 
boundary of the accretion disk closer to the neutron star surface.  In effect, 
this would imply a narrowing down of the boundary layer separation.  
Consequently, the boundary layer luminosity is expected to be  much smaller in 
comparison to the static or slowly rotating neutron star case, and this can 
alter the X--ray emission spectra of LMXBs.

The effect of rotation of the neutron star on the accretion luminosities was 
considered by
Datta, Thampan \& Wiita (1995), using the `slow' rotation (but general 
relativistic) formalism due to Hartle \& Thorne (1968).  These authors found 
that rotation
always increases the disk luminosity, usually decreases the boundary layer
luminosity and reduces the rate of angular momentum evolution, and gave
quantitative estimates corresponding to realistic neutron star models.  An 
important parameter in this connection is the radius of the innermost stable 
circular orbit ($r_{orb}$). This quantity plays a central role in deciding the 
magnitude of the gravitational energy release, and hence the accretion 
luminosities.  The relevance of $r_{orb}$ was emphasized by 
Klu\'{z}niak \& Wagoner (1985) who pointed out that for non--magnetic 
accreting neutron stars, it is incorrect to make the general assumption that 
the accretion disk will be separated from the neutron star surface by a thin 
boundary layer. The boundary layer separation  will depend on whether the 
equation of state (EOS) of neutron star matter is stiff or soft. For rapidly 
rotating neutron stars, Cook, Shapiro \& Teukolsky (1994) calculated the 
marginally stable circular orbits for application to angular momentum evolution of isolated 
neutron stars. 

Accretion onto a rapidly rotating neutron star  can bring in several 
interesting features. LMXBs are likely to accrete material whose total 
mass can be a substantial  fraction of the neutron star mass (\greq $0.1$  
\msun). This can severely reduce the magnitude of the boundary layer 
luminosity (King 1995).  Another  important question is whether or not  
the accreting neutron star will be disrupted once it reaches equilibrium  
rotation rate with further arrival of the accreted plasma. Recently,  
Bisnovatyi--Kogan (1993) has given a self--consistent analytical solution for 
an  
accretion disk structure around a rapidly rotating non--magnetized neutron star,  
using rigidly rotating polytropic   model. This work also  
gives a simple recipe for estimating the accretion luminosities based on  
accreting black--hole analogy.
In the present paper, we do not address the question of the disk structure or 
instability at  equilibrium rotation rates, but examine  how a rapid rotation 
rate of the neutron star will affect the 
boundary layer separation and reorder the contribution to the total accretion  
luminosity due to the disk and the boundary layer. The structure of 
rotating neutron stars in general relativity are 
calculated using a numerical code developed by us. This code is based on the
Komatsu, Eriguchi \& Hachisu (1989) formalism, as modified by Cook, 
Shapiro \& Teukolsky (1994) to incorporate realistic neutron star equations
of state.  This formalism is fully general  relativistic and  is amenable 
to a self--consistent numerical treatment, employing a Newton--Raphson type 
iterative 
scheme.  We find that rotation increases the disk luminosity and decreases 
the boundary layer luminosity, and for rotation rates near the centrifugal
mass shed limit, the total luminosity has contribution only from the 
extended disk.  Furthermore, for such maximal rotation rates, we find that
much before the maximum stable gravitational mass configuration is reached, 
there exists a 
limiting central density of the neutron star for which particles in the
innermost stable circular orbit will be more tightly bound than those at
the surface of the rotating neutron star. We also examine the possible
modifications in the angular velocity profile of the accreted material
in Keplerian orbit brought on by rapid rotation of the neutron star.

The format of this paper is arranged as follows.  Section 2 gives the 
formalism and the basic equations to be solved.  Section 3 outlines the
calculation of rotating neutron star models.
The Keplerian angular velocity profiles are described in section 4.  The 
results of our calculations are summarized in section 5 and a discussion
given in section 6.

\section{Accretion luminosities for a rotating space--time}

The space--time around a rotating neutron star can be described in 
quasi--isotropic coordinates, as a generalization of Bardeen's metric 
(Bardeen 1970):

\begin{eqnarray}
ds^2& = & - e^{\gamma+\rho} dt^2 + e^{2 \alpha}(dr^2+r^2d\theta^2) + 
\nonumber \\
& & ~~~~~~~~~~~~~~~~~~~~~~~~~~~~
e^{\gamma-\rho}r^2sin^2\theta (d\phi-\omega dt)^2 
\end{eqnarray}

\noindent where the metric potentials $\gamma$, $\rho$, $\alpha$, and the 
angular velocity of the stellar fluid relative to the local inertial frame 
($\omega$) are all functions of the quasi--isotropic radial coordinate ($r$) 
and the polar angle ($\theta$). We use here geometric units: $c=1=G$. 
Since the metric is stationary and axisymmetric, the energy and angular 
momentum are constants of motion.  Therefore, the specific energy $\tilde{E}$ 
(in units of the rest energy $m_{0}c^2$, where $m_{0}$ is the rest mass of the accreted 
particle) and the specific angular momentum $l$ (in units of $m_{0}c$) 
can be identified as $-p_{0}$ and $p_{3}$ respectively, where, $p_{\mu}$ 
($\mu = 0, 1, 2, 3$), stands for the four--momentum of the particle.  
From the condition $p_{\mu}p^{\mu} = -1$, we have the equations of motion  
of the particle (confined to the equatorial plane) in this gravitational field 
as 

\begin{eqnarray}
\dot{t} & = & \frac{dt}{d\tau} =  p^0  = e^{-(\gamma + \rho)} (\tilde{E} - 
\omega l) \\
\dot{\phi} & = & \frac{d\phi}{d\tau} =  p^3 = \Omega p^0 = 
e^{-(\gamma+\rho)} \omega (\tilde{E} - \omega l) + 
\frac{l}{r^2 e^{(\gamma-\rho)}} \\
\dot{r}^2 & \equiv & e^{2\alpha + \gamma + \rho}
 \left(\frac{dr}{d\tau}\right)^2 = 
\tilde{E}^{2} - \tilde{V}^2 .
\end{eqnarray}

\noindent Here $\Omega$ is the angular velocity of the star as seen by a distant
 observer, $d\tau$ is the proper time and $\tilde{V}$ is the effective
potential given by

\begin{eqnarray}
\tilde{V}^2 & = & e^{\gamma + \rho}
\left[1 + \frac{l^2/r^2} {e^{\gamma - \rho}}\right] 
+ 2\omega\tilde{E}l - \omega^{2}l^2.
\end{eqnarray}

\noindent The conditions for circular orbits, extremum of energy and minimum 
of energy are respectively: 

\begin{eqnarray}
\tilde{E}^2 & = & \tilde{V}^2 \\
\tilde{V}_{,r} & = & 0 \\
\tilde{V}_{,rr} & > & 0.
\end{eqnarray}

\noindent For marginally stable orbits,

\begin{eqnarray}
\tilde{V}_{,rr} & = & 0.
\end{eqnarray}

\noindent In our notation, a comma followed by one `$r$' represents a first order 
partial derivative with respect to $r$ and so on, and a tilde over a variable 
represents the corresponding dimensionless quantity.

From the expression for the effective potential and the conditions (6), (7) 
and (9), one obtains three equations in three unknowns, namely, $r$,
$\tilde{E}$, and $l$.  In principle, if analytical expressions for 
$e^{\gamma + \rho}$, $e^{2\alpha}$, $e^{\gamma - \rho}$ and $\omega$ are known, 
it would be a straightforward exercise to solve these equations to obtain $r$, 
$\tilde{E}$, and $l$. In practice, however, this is not so, and the 
solutions for the metric coefficients $e^{\gamma + \rho}$, $e^{2\alpha}$, 
$e^{\gamma - \rho}$, and $\omega$ have to be  obtained as arrays of numbers 
for various values of 
$r$ and $\theta$ using a numerical treatment. Furthermore, the condition (9) 
will introduce second order derivatives of $\gamma$, $\rho$, and $\omega$, 
which means that care has to be exercised in ensuring the numerical accuracies 
of the quantities calculated.  For this purpose, it is convenient to express 
$\tilde{E}$ and $l$ in terms of the physical velocity $\tilde{v}$ 

\begin{eqnarray}
\tilde{v} & = & (\Omega - \omega) r e^{-\rho} sin\theta 
\end{eqnarray}

\noindent of the stellar matter with respect to a locally nonrotating observer 
(see Bardeen 1972).

This gives the following expressions:

\begin{eqnarray}
\tilde{E}-\omega l & = & \frac{e^{(\gamma+\rho)/2}}
{\sqrt{1 - \tilde{v}^2}}\\
l & = & \frac{\tilde{v} r e^{(\gamma-\rho)/2}}
{\sqrt{1 - \tilde{v}^2}}.
\end{eqnarray}

\noindent Equations (11) and (12) can be recognized as the condition
 for circular orbits.  Conditions (7) and (9) yield respectively,

\begin{eqnarray}
 \tilde{v}  =  \frac{
e^{-\rho} r^{2} \omega_{,r}  \pm  
[e^{-2\rho}r^4\omega_{,r}^{2} + 2r(\gamma_{,r} + \rho_{,r}) +
r^2(\gamma_{,r}^2-\rho_{,r}^2)]^{1/2}}
{2 + r(\gamma_{,r} - \rho_{,r})} 
\end{eqnarray}

\begin{eqnarray}
\tilde{V}_{,rr} & \equiv & 2\left[\frac{r}{4}(\rho^2_{,r}-\gamma^2_{,r}) - 
\frac{1}{2}
 e^{-2\rho}\omega_{,r}^{2}r^3 - \rho_{,r} + \frac{1}{r}\right]\tilde{v}^2 
\nonumber \\
& &  + [2 + r(\gamma_{,r} - \rho_{,r})]\tilde{v}\tilde{v}_{,r} 
 - e^{-\rho}\omega_{,r} r \tilde{v} 
\nonumber \\
& & + \frac{r}{2}(\gamma^2_{,r} - \rho^2_{,r}) - e^{-\rho} r^2 \omega_{,r} 
\tilde{v}_{,r} = 0
\end{eqnarray}

\noindent where we have made use of Eq. (13) and its derivative with respect
to r in order to eliminate the second order derivatives in Eq. (14). 
The zero of $V_{,rr}$ will give the innermost stable circular orbit radius
($r_{orb}$) and the corresponding $\tilde{v}$ will
yield $\tilde{E}$ and $l$.  In equation (13), the positive sign refers 
to the co--rotating particles and the negative sign to the counter--rotating
particles.  In this  study we have considered only the co--rotation case.

Depending on the EOS and the central density, neutron stars can have radii 
greater than or less than $r_{orb}$ (Datta, Thampan \& Wiita 1995).  The 
accretion luminosities will, of course, be different for these two cases  
(Klu\'{z}niak and Wagoner 1985; Sunyaev and Shakura 1986; Datta, Thampan 
and Wiita 1995). These quantities can be calculated as follows:

\noindent Case (a): Radius of the star ($R$) greater than $r_{orb}$.

If an accretion disk were to form around a relatively large neutron star, the
ingress of a particle of rest mass $m_0$ from infinity to the disk boundary
(which will be at the stellar surface) will release an amount of energy given
by:

\begin{eqnarray}
E_{D} & = & m_0\{1 - \tilde{E}_{K}(r=R)\}
\end{eqnarray}

\noindent where $\tilde{E}_{K}(r=R)$ is the specific energy of the particle in
Keplerian orbit at the surface obtained by solving equation (13) to obtain
$\tilde{v}_{K}=\tilde{v}$ and solving equations (11) and (12) with $r=R$ and 
$\tilde{v}=\tilde{v}_{K}$ to obtain $l_{K}$ and $\tilde{E}_{K}(r=R)$.

The energy loss in the boundary layer (a very narrow gap near the neutron
 star surface) will 
be

\begin{eqnarray}
E_{BL} & = & m_0\{\tilde{E}_{K}(r=R) - \tilde{E}_0\}
\end{eqnarray}

\noindent where $\tilde{E}_0$ is the energy of the particle ``at rest'' on
the stellar surface (the particle will be moving with the velocity
$\tilde{v}=\tilde{v}_\ast$ of the stellar fluid at the surface, where
$\tilde{v}_\ast$ is obtained by substituting into equation (10) all the
relevant parameters for $r=R$) and is calculated by solving equations
(11) and (12) for $\tilde{E}$ at $r=R$ and $\tilde{v}=\tilde{v}_\ast$.

\noindent Case (b): Radius of the star (R) smaller than radius of 
$r_{orb}$.

In this case, the accretion disk will extend inward to a radius corresponding
to $r=r_{orb}$. The energy released in the disk as the particle
comes in from infinity to the innermost stable circular orbit will be

\begin{eqnarray}
E_{D} & = & m_0\{1 - \tilde{E}_{orb}\}.
\end{eqnarray}

\noindent The energy released in the boundary layer will be

\begin{eqnarray}
E_{BL} & = & m_0\{\tilde{E}_{orb} - \tilde{E}_0\}
\end{eqnarray}

\noindent where $\tilde{E}_{orb}$ is the energy of the particle in innermost
stable circular orbit, calculated 
by finding the $r=r_{orb}$ at which equation (14) is satisfied and then
solving equations (11), (12) and (13) for this $r$ to yield $\tilde{E}_{orb}$.
 The energy $\tilde{E}_0$ of the particle on the stellar surface is calculated
as described in the previous case.

\section{Angular velocity profiles}

For slow rotation of the neutron star, the angular velocity of the accreted
material in Keplerian orbit around it, $\Omega(r)$, will have a profile 
that has a maximum that is located outside the neutron star surface.  For rapid
rotation rates of the star (corresponding to angular velocity close to
the Keplerian value at the surface), a second type of profile 
for $\Omega(r)$
is also possible, in which $\Omega(r)$ exhibits no maximum but increases 
monotonically all the way to the surface of the neutron star.  In such a
situation, the accretion torque on the neutron star will not be purely
advective.  It will become possible for the viscous torque to transport
angular momentum outwards at all radii.  This can lead to interesting 
accretion scenarios.

The Keplerian angular velocity $\Omega_K$ of a particle in an orbit around
the rotating neutron star is defined as:

\begin{eqnarray}
\Omega_K(r) & = & e^{\rho(r)} \frac{\tilde{v}(r)}{r} + \omega(r)
\end{eqnarray}

\noindent where $\tilde{v}$ is as given in equation (12).  The Keplerian 
angular velocity of the particle in an orbit at the surface of the neutron
star puts a firm upper
bound on the angular velocity the star can attain (Friedman, Ipser and Parker
1986) and hence the boundary layer luminosity when the star attains this
maximum $\Omega$ should be zero (Sunyaev and Shakura 1986).  

\section{Rapidly rotating neutron star models in general relativity}

Rapidly rotating neutron star structure in general relativity for realistic
neutron star EOS have been reported by Friedman, Ipser \& Parker (1986).
Their numerical code is based on the programmes developed by Butterworth 
\& Ipser (1976). Previous models of rapidly rotating neutron stars have
been based on incompressible fluids and polytropic models (Bonazzola \&
Schneider 1974; Butterworth 1976). Komatsu, Eriguchi \& Hachisu (1989)  
have generalized the Newtonian self--consistent field method to a 
general relativistic case to obtain structures of rapidly rotating stars,
again using the polytropic model. This technique was modified by Cook,
Shapiro \& Teukolsky (1994) for realistic neutron star EOS for 
purpose of studying quasi--stationary evolution of isolated neutron
stars.  A variant of this approach based on spectral methods was developed 
by Bonazzola et al. (1993).

In this investigation, we have calculated the structure of 
rapidly rotating neutron stars in general relativity using a numerical
code developed by the present authors, which is based on the method due
to Komatsu, Eriguchi \& Hachisu (1989), as modified by Cook, Shapiro
\& Teukolsky (1994) so as to incorporate realistic neutron star EOS.
The results of our code agree with the published results of Friedman, 
Ipser \& Parker (1986), and results using the code of Stergioulas \& Friedman 
(1995) to less than 1\%. Also, wherever a comparison was possible, our
results agreed with those reported in Cook, Shapiro \& Teukolsky (1994) 
to a similar degree of accuracy. 
For non--rotating equilibrium models, we found our results to be within 
 0.3\% of the published results of Arnett and Bowers (1977).  We have 
constructed numerically various sequences of neutron stars, starting from 
the static limit all the way upto the rotation rate corresponding to the 
centrifugal
 mass  shed limit. The latter limit corresponds to the maximum $\Omega$ 
for which  centrifugal forces are able to  balance the inward gravitational 
force. Any further increase in $\Omega$ will lead to disruption of the star. 
The general relativistic expression for this limit can be found in 
Cook, Shapiro \& Teukolsky (1994).

\begin{table*}
 \centering
 \begin{minipage}{170mm}
  \caption{
The values of the neutron star gravitational mass 
$M_G$, disk luminosity $E_D$, boundary layer luminosity $E_{BL}$, 
the boundary layer separation (i.e, the height above stellar surface, 
where the innermost stable circular orbit is located) $h^+$ and the 
Keplerian angular velocity $\Omega_{K}$ for two values of neutron star
rotation rates $\Omega=0$ and $\Omega=\Omega_{ms}$ and chosen values of
central density $\rho_c$ -- see text for details.  The numbers 
following the letter $E$ in column 9 stand for powers of ten.  $\rho_c$
is in units of $10^{14}$ \gcc; $\Omega_{ms}$ and $\Omega_K$ are in units of
$10^4 ~rad ~s^{-1}$.}
 \begin{tabular}{crccccccrcccc}                          \hline\hline
 & & & & & & & & & & & & \\
 \multicolumn{1}{c}{EOS}           & \multicolumn{1}{c}{$\rho_c$}%
& \multicolumn{1}{c}{$\Omega_{ms}$}%
& \multicolumn{2}{c}{$M$}      & 
 \multicolumn{2}{c}{$E_D$}         & \multicolumn{2}{c}{$E_{BL}$}    & 
 \multicolumn{2}{c}{$h^+$}         & \multicolumn{2}{c}{$\Omega_K$} \\ 
 & & & & & & & & & & & & \\
 \multicolumn{1}{c}{}              & 
\multicolumn{1}{c}{}     & 
 \multicolumn{1}{c}{}  & 
\multicolumn{2}{c}{(\msun)}    & 
 \multicolumn{2}{c}{($m_0~c^2$)}     & \multicolumn{2}{c}{($m_0~c^2$)}& 
 \multicolumn{2}{c}{(km)}           
 & \multicolumn{2}{c}{}           \\   \cline{4-13}
 & & & & & & & & & & & & \\
 \multicolumn{1}{c}{}              & \multicolumn{1}{c}{}         & 
 \multicolumn{1}{c}{}              & 
 \multicolumn{1}{c}{$\Omega=0$}    & \multicolumn{1}{c}{$\Omega=\Omega_{ms}$} &
 \multicolumn{1}{c}{$\Omega=0$}    & \multicolumn{1}{c}{$\Omega=\Omega_{ms}$} &
 \multicolumn{1}{c}{$\Omega=0$}    & \multicolumn{1}{c}{$\Omega=\Omega_{ms}$} &
 \multicolumn{1}{c}{$\Omega=0$}    & \multicolumn{1}{c}{$\Omega=\Omega_{ms}$} &
 \multicolumn{1}{c}{$\Omega=0$}    & \multicolumn{1}{c}{$\Omega=\Omega_{ms}$} \\
\hline
 & & & & & & & & & & & & \\

(A)  & 26.630 & 1.105 & 1.183 & 1.400  & 0.057 & 0.073 & 0.177 &    1.578$E-4$ & 2.052 & 0.000 & 1.167 & 1.106 \\
     & $^\ast$39.121& 1.370 & 1.357 & 1.585  & 0.057 & 0.084 & 0.246 &    1.141$E-4$ & 4.239 & 0.000 & 1.018 & 1.370\\
     & 43.000 & 1.434 & 1.381 & 1.608  & 0.057 & 0.085 & 0.261 &   $-$9.000$E-6$ & 4.616 & 0.223 & 1.000 & 1.387\\ 
 & & & & & & & & & & & & \\
(B)  & 10.670 & 0.718 & 1.133 & 1.400  & 0.056 & 0.058 & 0.110 &   4.970$E-5$ &  ---  &  ---  & 1.065 & 0.718\\
     & 13.730 & 0.847 & 1.400 & 1.723  & 0.057 & 0.071 & 0.160 &   6.780$E-5$ & 1.719 & 0.000 & 0.986 & 0.847\\
     & $^\ast$22.600 & 1.062 & 1.694 & 2.034  & 0.057 & 0.084 & 0.237 &   1.866$E-4$ & 5.057 & 0.000 & 0.815 & 1.063\\
     & 26.000 & 1.119 & 1.730 & 2.060  & 0.057 & 0.086 & 0.254 &   1.160$E-5$ & 5.619 & 0.286 & 0.798 & 1.082\\
     & 35.270 & 1.240 & 1.757 & 2.060  & 0.057 & 0.088 & 0.284 &  $-$9.098$E-4$ & 6.403 & 0.721 & 0.786 & 1.133\\
 & & & & & & & & & & & & \\
(C)  &  6.363 & 0.609 & 1.088 & 1.400  & 0.053 & 0.053 & 0.090 &   3.880$E-5$ &  ---  &  ---  & 0.900 & 0.609\\
     &  7.413 & 0.689 & 1.400 & 1.818  & 0.057 & 0.065 & 0.128 &   1.058$E-4$ & 0.137 &  ---  & 0.985 & 0.689\\
     & $^\ast$10.610 & 0.845 & 1.961 & 2.493  & 0.057 & 0.084 & 0.217 &   2.199$E-4$ & 5.140 & 0.000 & 0.704 & 0.845 \\
     & 19.500 & 1.042 & 2.279 & 2.769  & 0.057 & 0.094 & 0.303 &  $-$3.196$E-3$ & 8.799 & 1.408 & 0.606 & 0.905\\
     & 21.750 & 1.070 & 2.284 & 2.757  & 0.057 & 0.094 & 0.311 &  $-$3.877$E-3$ & 9.020 & 1.508 & 0.605 & 0.918\\
 & & & & & & & & & & & & \\
(D)  & 8.757 & 0.700 & 1.106 & 1.400  & 0.056 & 0.057 & 0.104 &   5.593$E-5$  & ---   &  ---  & 1.033 & 0.700\\
     & 10.430 & 0.803 & 1.400 & 1.780  & 0.057 & 0.070 & 0.150 &   7.670$E-5$  & 1.297 & 0.000 & 0.986 & 0.803\\
     & $^\ast$13.530 & 0.953 & 1.788 & 2.239  & 0.057 & 0.084 & 0.223 &   1.055$E-4$  & 4.896 & 0.000 & 0.772 & 0.954\\
     & 22.270 & 1.217 & 2.160 & 2.604  & 0.057 & 0.097 & 0.331 &  $-$6.517$E-3$  & 8.946 & 1.640 & 0.640 & 1.013\\
 & & & & & & & & & & & & \\
(E)  &  3.556 & 0.465 & 1.059 & 1.400  & 0.047 & 0.047 & 0.069 &   1.412$E-4$ &  ---  &  ---  & 0.692 & 0.466\\
     &  4.064 & 0.520 & 1.400 & 1.887  & 0.055 & 0.057 & 0.097 &   7.938$E-5$ &  ---  &  ---  & 0.761 & 0.520\\
     &  $^\ast$6.996 & 0.674 & 2.338 & 3.043  & 0.057 & 0.083 & 0.207 &   1.836$E-4$ & 5.641 & 0.000 & 0.591 & 0.674\\
     & 11.000 & 0.762 & 2.572 & 3.221  & 0.057 & 0.089 & 0.254 &  $-$4.476$E-4$ & 8.341 & 0.865 & 0.537 & 0.711\\
     & 13.394 & 0.794 & 2.589 & 3.200  & 0.057 & 0.089 & 0.266 &  $-$8.943$E-4$ & 8.841 & 1.038 & 0.534 & 0.730\\
 & & & & & & & & & & & & \\
(F)  &  9.136 & 0.697 & 1.106 & 1.400  & 0.056 & 0.057 & 0.104 &  1.149$E-4$ &  ---  &  ---  & 1.030 & 0.697\\
     & 11.900 & 0.812 & 1.398 & 1.760  & 0.057 & 0.070 & 0.152 &  1.012$E-4$ & 1.382 & 0.000 & 0.988 & 0.812\\
     & $^\ast$20.493 & 1.019 & 1.732 & 2.109  & 0.057 & 0.084 & 0.233 &  9.228$E-5$ & 5.031 & 0.000 & 0.797 & 1.019\\
     & 26.000 & 1.100 & 1.780 & 2.135  & 0.057 & 0.086 & 0.257 & $-$1.487$E-4$ & 5.837 & 0.380 & 0.776 & 1.053\\
     & 30.890 & 1.155 & 1.788 & 2.126  & 0.057 & 0.087 & 0.270 & $-$5.163$E-4$ & 6.199 & 0.568 & 0.772 & 1.080\\
 & & & & & & & & & & & & \\
\hline
\end{tabular}
\end{minipage}
\end{table*}

The structure of neutron stars depends  sensitively on the EOS at high
densities.  Although the main composition of degenerate matter at 
densities higher than nuclear matter density is expected to be dominated
by neutrons, significant admixtures of other elementary particles (such
as pions, kaons and hyperons) are not ruled out.  A persistent problem in
determining the EOS for neutron star interior is what to choose for the
interaction potential among the constituent particles, for which reliable
experimental data are not available.  All calculations involve either
extrapolations from known nuclear matter properties or plausible field
theoretical approaches using mean field approximation. Another related but 
unresolved problem is: what is an adequate many--body technique to
estimate the higher order correlation terms in expressions for the pressure.
In this paper we do not address these problems, but choose, for illustrating
the results of the present study, six EOS models based on representative 
neutron star matter interaction models.  This is expected to provide a 
broad realistic set of conclusions.  A brief description of the EOS models
used here is given below.

(A) {\it Pandharipande (1971) (hyperonic matter)}:
One of the early attempts to derive nuclear EOS with admixture of hyperons is 
due to Pandharipande (1971),
who assumed the hyperonic potentials to be similar to the nucleon--nucleon 
potentials, but altered suitably to represent the different isospin states.
The many--body method adopted is based on the variational approach of 
Jastrow (1955).  The two body wave function was taken as satisfying a
simplified form of the Bethe--Goldstone equation, in which terms representing
the Pauli exclusion principle were omitted but simulated by imposing a `healing'
constraint on the wave function.

(B) {\it Bethe-Johnson}:  
Bethe \& Johnson (1974) devised phenomenological potentials for
nucleon--nucleon interaction that have realistic short--range
behaviour.  These authors then used the lowest order constrained variational
method to calculate the EOS of neutron star matter.  The work of Bethe \& 
Johnson (1974) consists of two different parts: (a) determination of the
EOS for pure neutron gas and (b) derivation of a hyperonic equation of
state.  For purpose of illustration here, we have chosen their EOS model 
V (neutron matter).

(C) {\it Walecka (1974) (neutrons)}: The EOS model of Walecka (1974) corresponds
to pure neutron matter, and is based on a mean--field theory with exchange
of scalar and (isoscalar) vector mesons representing the nuclear interaction.

(D) {\it Wiringa, Fiks} \& {\it Fabrocini (1988) (UV14 + UVII)}: These authors gave a 
model of EOS for dense nuclear and neutron matter which includes three--nucleon
interactions.  This is a non--relativistic approach based on the variational 
method.   The three--body potential includes long--range repulsive parts
that are adjusted to give light nuclei binding energies and nuclear matter 
saturation properties. The authors have given three models.  We consider here
their model UV14+UVII for beta--stable case: neutrons, protons, electrons and
muons.

(E) {\it Sahu, Basu} \& {\it Datta (1993)}: gave a field theoretical EOS for neutron--rich
matter in beta equilibrium based on the chiral sigma model.  The model includes
an isoscalar vector field generated dynamically and reproduces the empirical
values of the nuclear matter saturation density and binding energy and also 
the isospin symmetry coefficient for asymmetric nuclear matter.  The energy
per nucleon of nuclear matter according to Sahu, Basu \& Datta (1993) is in
very good agreement, up to about four times the equilibrium nuclear matter
density, with estimates inferred from heavy--ion collison experimental data.

(F) {\it Baldo, Bombaci} \& {\it Burgio (1997)}: have recently given a microscopic
EOS for asymmetric nuclear matter, derived from the Brueckner--Bethe--Goldstone
 many--body theory with explicit three--body terms. The three--body
force parameters are adjusted to give a reasonable saturation point for
nuclear matter.

\begin{figure*}
\hspace{-2.5cm}
{\mbox{\psboxto(20cm;10cm){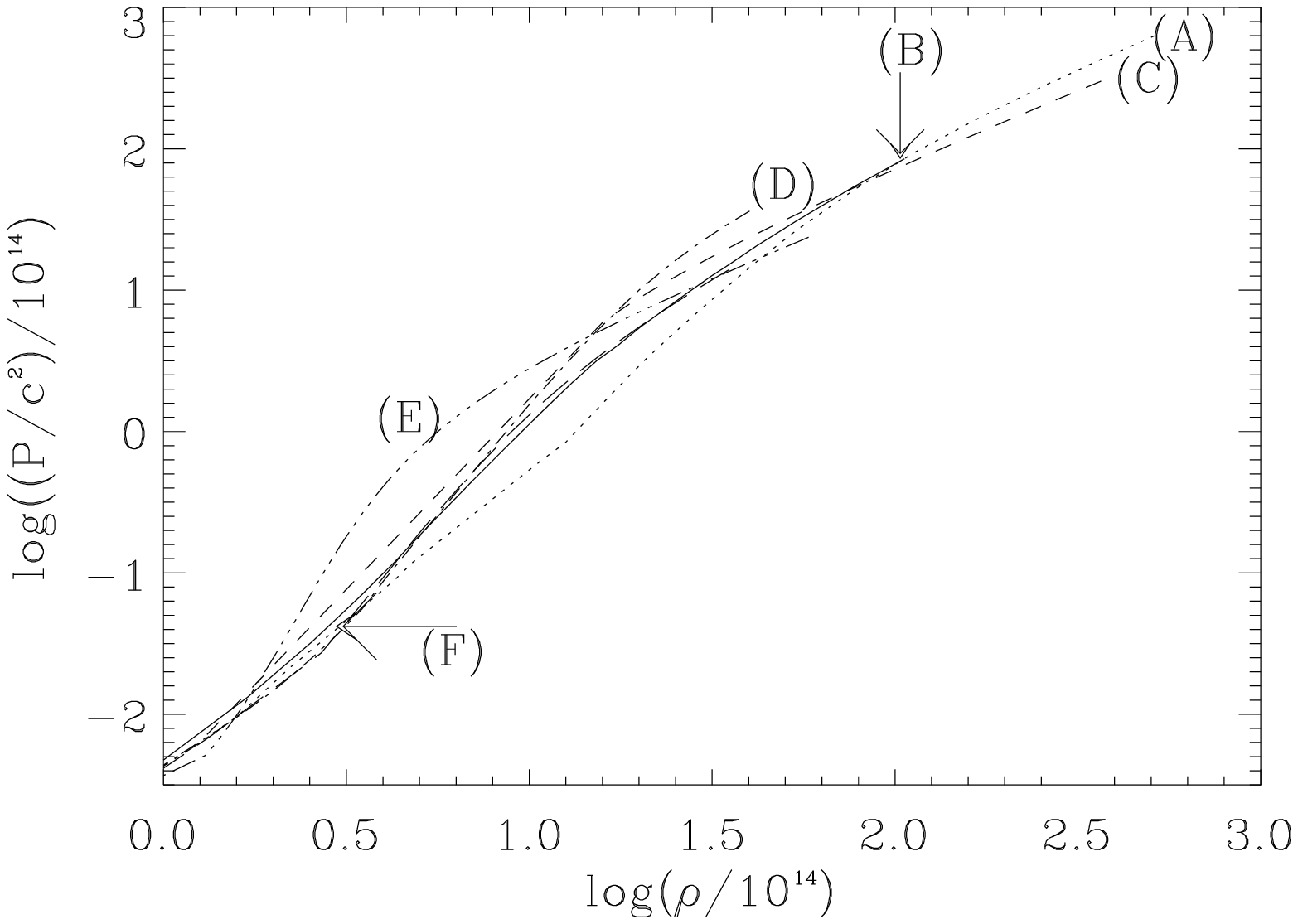}}}
 \caption{Pressure as a function of density for the EOS models (A--F).}
\end{figure*}

The pressure--density relationship of these above EOS models is illustrated
in Fig 1.  Of the above EOS, model (A) is a relatively soft EOS, models (B), 
(D) and (F) are roughly intermediate in stiffness, whereas models (C) and 
(E) are rather stiff EOS. It may be recalled that a stiffer EOS will give a 
higher value for the non--rotating neutron star maximum mass and also
a higher value for the corresponding radius. The composite EOS for the 
entire span of neutron star densities was constructed by joining the
selected high density EOS to that of Negele \& Vautherin (1973) for the
density range ($10^{14} - 5\times10^{10}$) \gcc, Baym, Pethick \& 
Sutherland (1971) for densities down to $\sim 10^3$ \gcc and Feynman,
Metropolis \& Teller (1949) for densities less than $10^3$ \gcc.

We found that while the results for EOS's (B), (C), (E) and (F) were 
straightforward,
EOS models (A) and  (D) presented the peculiarity of having the 
maximum $\Omega$ model as the maximum stable mass model (in agreement with the 
result of Cook, Shapiro \& Teukolsky 1994) suggesting that EOS's (A) and (D) 
belong 
to the Class I EOS and EOS models (B), (C), (E) and (F) to Class II as 
classified in 
Stergioulas \& Friedman (1995).

\section{Results}

All the calculated parameters depend on the central density ($\rho_c$) and 
rotation rate ($\Omega$) of the neutron star. In Table 1 we have summarized the 
functional dependences on $\rho_c$ of some important parameters 
in the calculation of luminosities.  In order to illustrate this dependence, 
we choose two limits of $\Omega$, namely, the non-rotating or static limit 
($\Omega=0$) and the centrifugal mass shed limit ($\Omega=\Omega_{ms}$),  
which are the two natural ends of a constant density sequence.  As expected,
the functional form of the dependence of structure parameters on central 
density for rotation rate at centrifugal mass shed limit ($\Omega=\Omega_{ms}$) 
is found to be qualitatively similar to that at the static limit and so
are not explicitly displayed here (see, for example Friedman, Ipser \& Parker 
1986).  The listed quantities in Table 1 are 
the values of the neutron star gravitational mass 
$M_G$, disk luminosity $E_D$, boundary layer luminosity $E_{BL}$, the boundary 
layer separation (i.e, the height above stellar surface, where the innermost 
stable circular orbit is located) $h^+$ and the Keplerian angular velocity 
$\Omega_{K}$.
 Generally speaking, $r_{orb}$ exhibits three characteristics: 
(a) $r_{orb}$ is non--existent, (b) $r_{orb} < R$, and (c) $r_{orb} > R$.  For 
the 
first two cases, the $r_{orb}$ is taken to be the Keplerian orbit radius at the
surface of the star.  The cases for which $r_{orb}$ is non--existent are 
differentiated in Table 1 by a dashed line under the column for $h^{+}$ and the cases 
for which $r_{orb} \le R$ are indicated by the entries for which $h^{+}$ is 
zero.  The central density at which $r_{orb}$ exactly equals $R$ in the 
rotating case (we call this the radius cross--over density), is indicated by 
an asterisk over the corresponding central density. For each EOS model, we 
choose five central densities corresponding to: (i) $M_G = 1.4$ \msun
at $\Omega=\Omega_{ms}$, (ii) $M_G = 1.4$ \msun at $\Omega=0$, (iii) the
radius cross--over density, at which $r_{orb}=R$ for $\Omega=\Omega_{ms}$,
(iv) $M_G = $ maximum mass at $\Omega=\Omega_{ms}$ and (v) $M_G = $ maximum
mass at $\Omega=0$. For neutron stars corresponding to the EOS models (A) and 
(D), we find that before the maximum stable mass is attained, the sequences become unstable to
radial perturbations, and hence the corresponding maximum $\Omega$ model can
be taken as the maximum mass stable model. Therefore, in Table 1 we have 
listed, for these EOS, parameters corresponding to the maximum $\Omega$ model 
instead of the maximum mass model.

From Table 1, it can be seen that typical increases in mass (to maintain the 
central density constant) from the static limit is (14--35)\%, with the larger 
changes corresponding to  the lower densities in the stiffer EOS. The 
corresponding 
increase in radius (see Fig. 4) lie in the range (28--44)\%.

Table 1 also shows that the disk luminosity $E_D$ increases for 
$\Omega=\Omega_{ms}$ with central density whereas, in the non--rotating case, 
it remains constant (except in the low central density regime for the stiff 
EOS for which the $r_{orb}$ is located at the surface of the star).  On the 
other hand, $E_{BL}$ is substantially higher (almost by a factor of 2) than 
$E_D$ in the static limit but becomes almost zero at the centrifugal mass shed 
limit 
(because of the fact that rotation rate at this limit is very nearly equal 
to the Kepler frequency of a particle in the innermost stable circular orbit at  
the surface). Interestingly, $E_{BL}$ is negative for densities
higher than the radius cross--over  density.

\begin{figure*}
\hspace{-2.5cm}
{\mbox{\psboxto(20cm;10cm){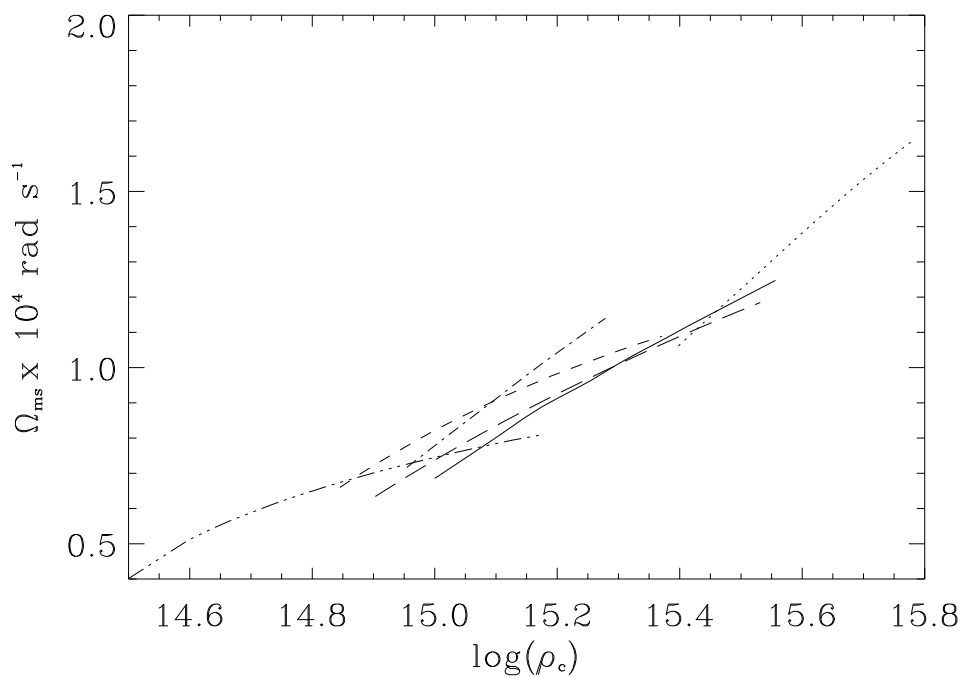}}}
 \caption{The neutron star rotation rate at the centrifugal mass shed
limit as a function of central density for various EOS models. The 
labels in Figs 2, 3, 5 \& 7 are as in Fig 1.}
\end{figure*}

In Fig. 2, we display the dependence of $\Omega_{ms}$ with $\rho_c$.
$\Omega_{ms}$ varies monotonically with $\rho_c$. Furthermore, $\Omega_{ms}$
seems to possess a marked dependence on the EOS; the softer the EOS, the
larger the value of $\Omega_{ms}$.  All curve labels in Figs 2, 3, 5 and 7
are as indicated in Fig 1.

\begin{figure*}
\hspace{-2.5cm}
{\mbox{\psboxto(20cm;10cm){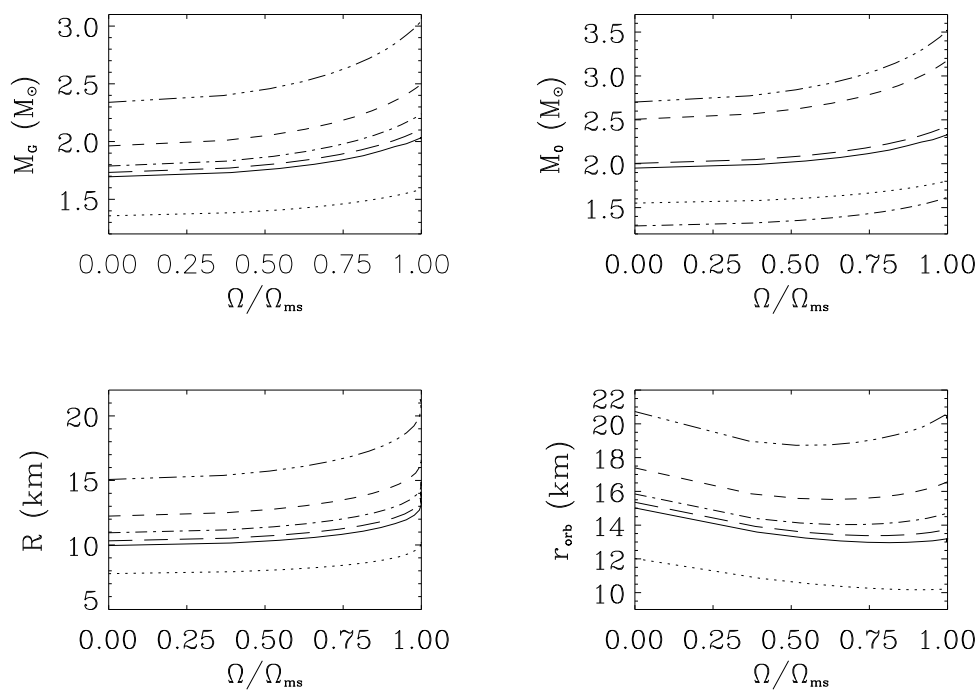}}}
 \caption{ Neutron star gravitational mass ($M_G$), baryonic mass 
($M_0$), radius ($R$), and the radius ($r_{orb}$) of the innermost
stable circular orbit as a function of the star's rotation rate for 
        various EOS.  The graphs correspond to the radius cross over
        central density for each EOS.
}
\end{figure*}

\noindent The variation of the gravitational mass $M_G$, baryonic mass $M_0$, 
stellar radius $R$, and the radius $r_{orb}$ of the innermost stable circular
orbit with respect to $\Omega$ for the radius cross--over density is illustrated 
in Fig. 3. From Fig 3, it can be seen that until $\Omega \sim 0.5\Omega_{ms}$, 
the structure parameters change slowly but for higher rotation rates ($\Omega$ 
\greq $0.6 \Omega_{ms}$), the rate of change is more pronounced. These changes 
are EOS dependent, being more substantial for stiffer EOS.

\begin{figure*}
\hspace{-1.0cm}
{\mbox{\psboxto(18cm;14cm){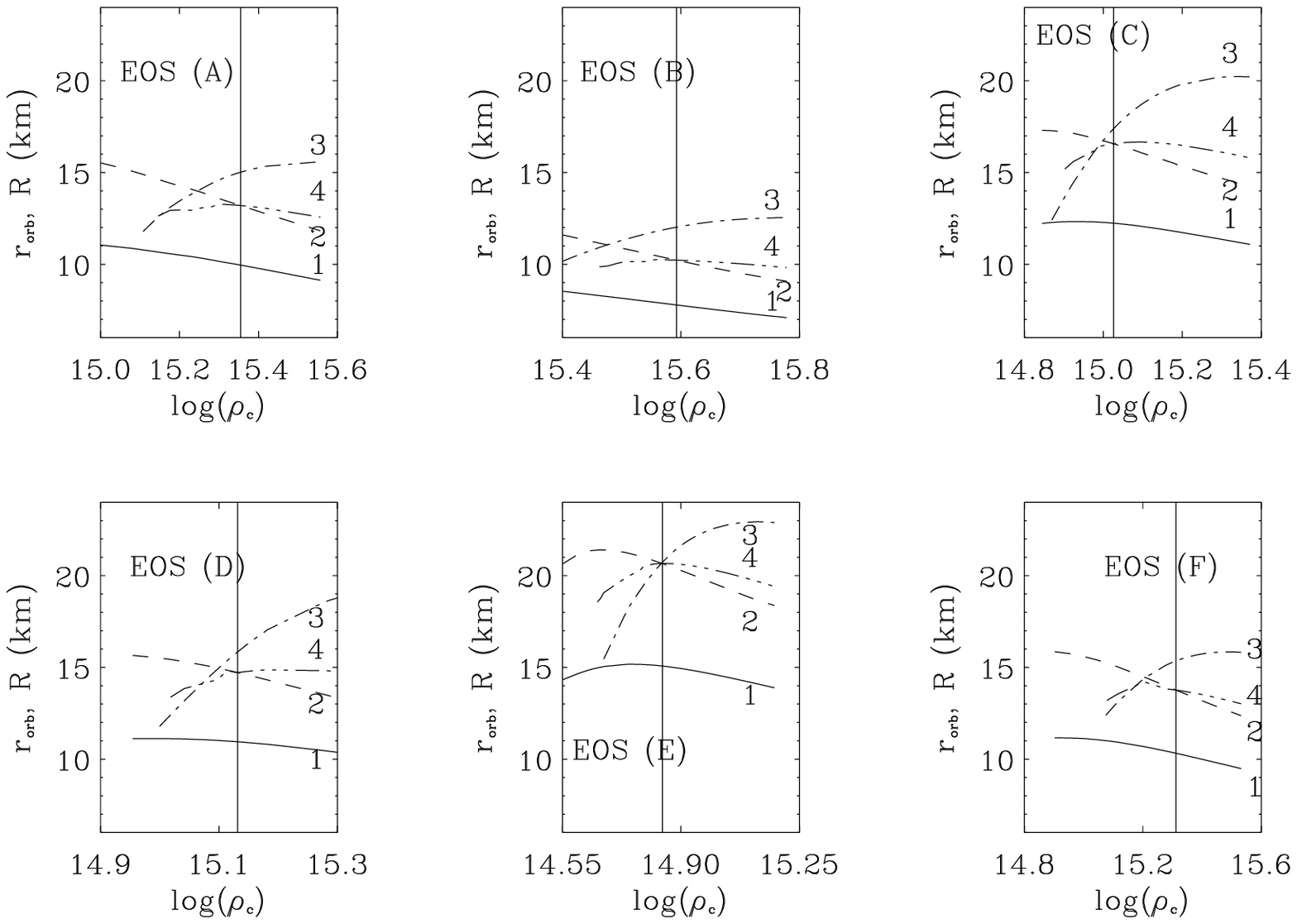}}}
 \caption{Neutron star radius ($R$) and radius ($r_{orb}$) of the 
innermost stable circular orbit as functions of central density for
two values of stellar rotation rate ($\Omega$). See text for the 
        details.
}
\end{figure*}

To illustrate how the boundary layer separation varies with $\rho_c$, $\Omega$ 
and also with the EOS, we give in Fig. 4, plots of $r_{orb}$ and $R$  versus 
$\rho_{c}$ for two cases of $\Omega$.  In Fig. 4, the six graphs corresponding 
to the different EOS models, display $R(\Omega=0)$ (curve 1), 
$R(\Omega=\Omega_{ms}$ (curve 2),  $r_{orb}(\Omega=0)$ (curve 3) and 
$r_{orb}(\Omega=\Omega_{ms})$ (curve 4).  It is immediately apparent from the 
plots that as $\Omega$ increases from $0$ to $\Omega_{ms}$ for a fixed 
$\rho_{c}$, R (curves 1 and 2) increases and $r_{orb}$ (curves 3 and 4) 
decreases.  Furthermore, it can be seen that in the static case $r_{orb}$  
is generally greater than $R$ for the whole range of central densities, the 
exception being for the stiff EOS (C) and (E) (in these latter cases, $r_{orb}$ 
is less than R for a few lower central density values), whereas at the 
centrifugal mass shed limit, $r_{orb}$  is greater than $R$  only for very high 
densities.   The intersection of curve (2) with curve (4) represents the 
cross--over density $\rho_c$ for which $r_{orb}=R$ at the centrifugal mass shed 
limit.  Since at the mass shed limit, the stellar rotation rate is almost equal 
to the angular velocity of a particle in Keplerian orbit at the surface of the 
star, the radius cross--over density represents the point at which the particle 
in the innermost stable circular orbit will start becoming more bound than a 
particle at the surface of the star.  In other words,  $E_{BL}$ will tend to   
zero at this central density and will become negative at some central density 
greater than this. This is indicated by the vertical line in the graph.

\begin{figure*}
\hspace{-2.5cm}
{\mbox{\psboxto(20cm;10cm){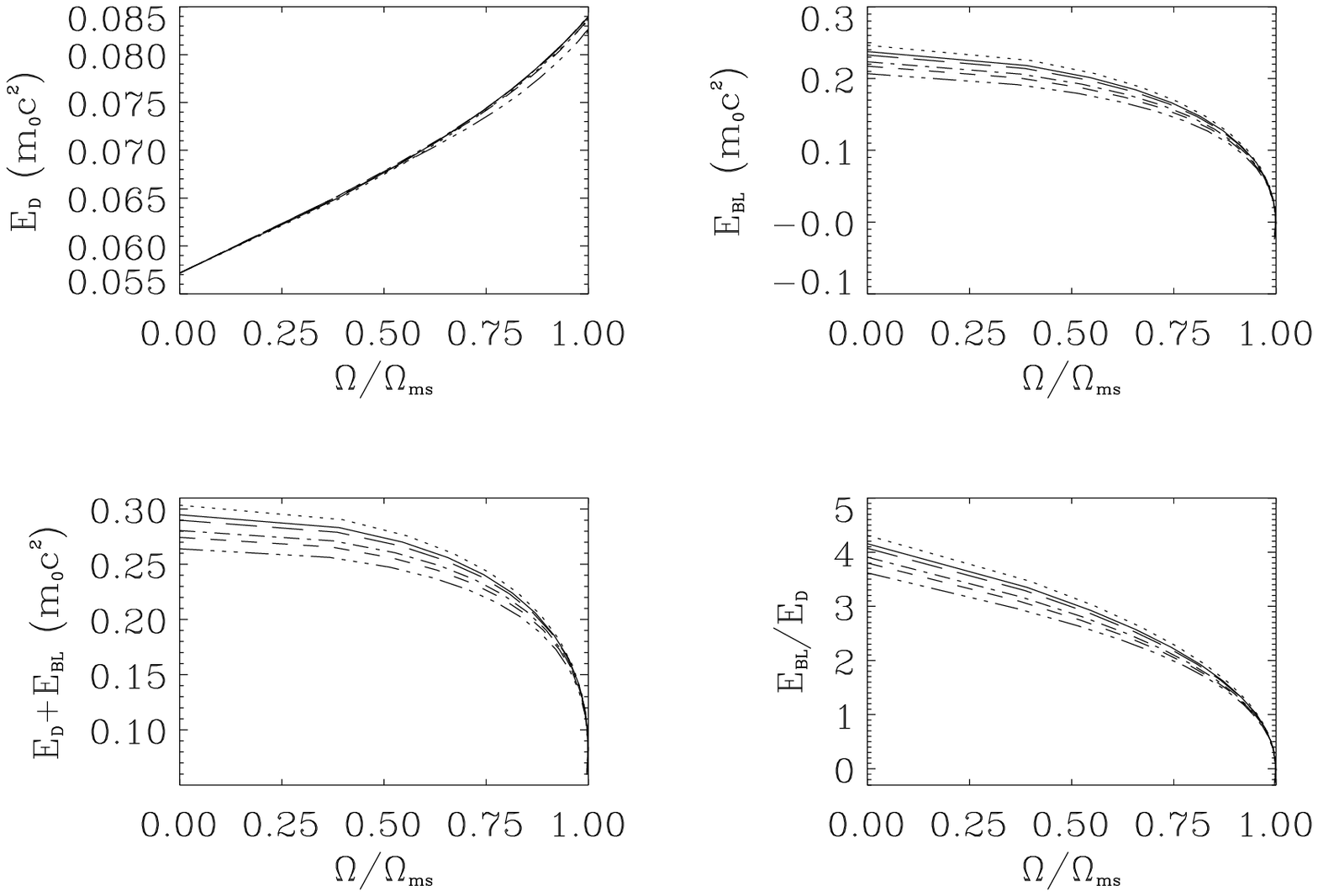}}}
 \caption{ The accretion luminosities in units of the rest mass of the 
        accreted particle as functions of neutron star rotation rate.
        The graphs correspond to the radius cross--over central 
        density for each EOS.
}
\end{figure*}

In Fig. 5, we plot $E_D$, $E_{BL}$, the total luminosity ($E_D+E_{BL}$) and
the ratio of $E_{BL}$ to $E_D$, all as functions of $\Omega/\Omega_{ms}$.  All 
these plots are for radius cross-over density for the various EOS.  Among the 
various EOS, $E_D$ differs most when $\Omega$ is large.  In contrast, $E_{BL}$ 
differs (among the various EOS), most for slow rotation rates. Therefore, the 
total luminosity  follows a variation similar to that of $E_{BL}$.

\begin{figure*}
\hspace{-1.0cm}
{\mbox{\psboxto(18cm;14cm){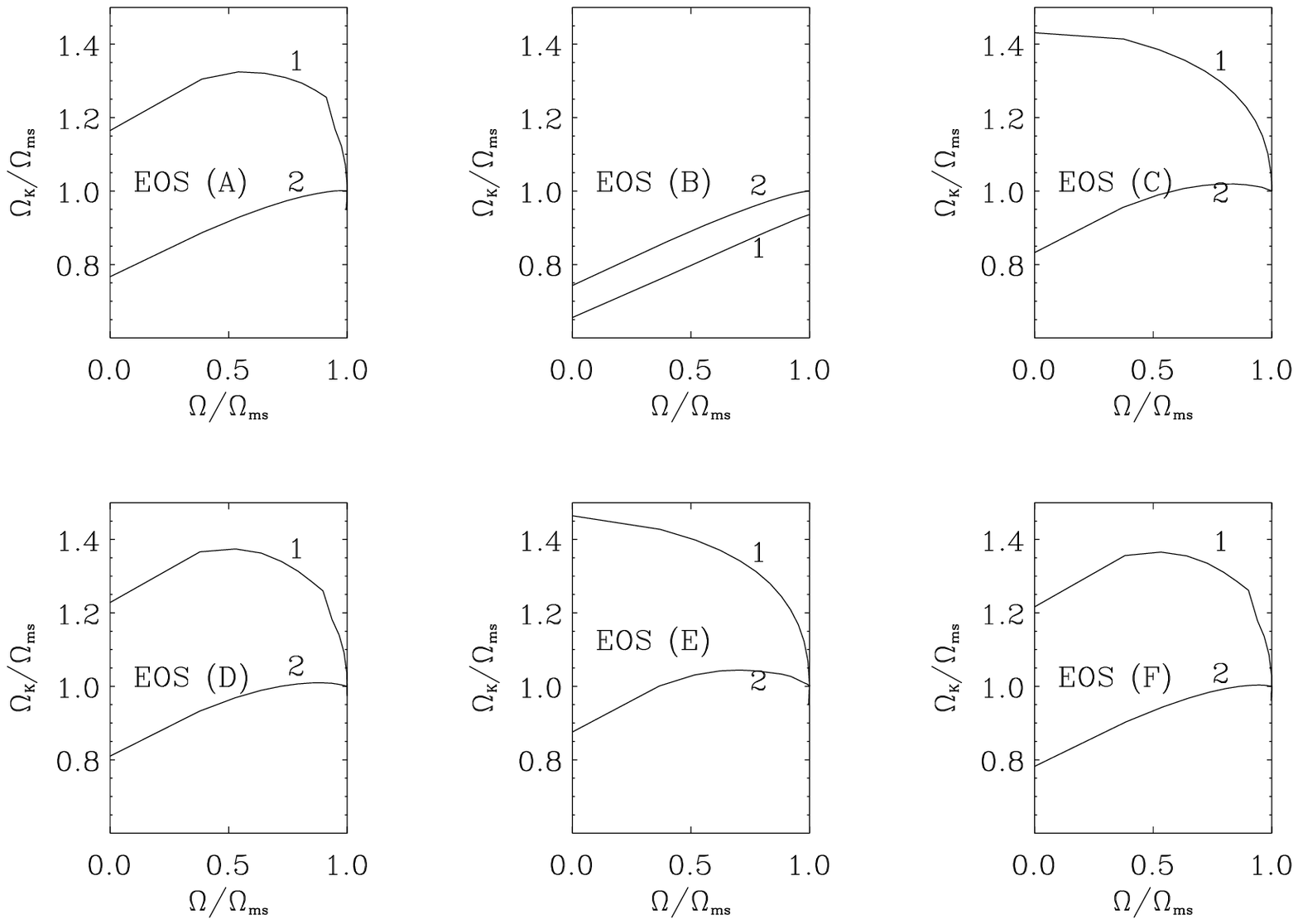}}}
 \caption{ 
 The Keplerian angular velocity of a particle in the innermost 
        stable circular orbit as a function of the neutron star
     rotation rate. Curve 1 refers to the central density corresponding
        to a gravitational mass of 1.4 \msun in the static limit, and
        curve 2 corresponds to that at the radius cross--over central
        density.
}
\end{figure*}

The dependence of $\Omega_K$ on $\Omega$ is shown in Fig 6.  For convenience  
of display and comparison, these are taken in units of $\Omega_{ms}$, and two 
chosen values of $\rho_{c}$. In Fig 6 curve 1 refers to $\rho_c$ corresponding
to $M_G=1.4$ \msun at static limit and curve 2 refers to $\rho_c$ for the 
radius cross--over density.  The qualitative differences in the behaviour of 
curve 1 for all the graphs is quite noticeable.  These differences arise
entirely due to the differences in the location of the innermost stable 
circular orbit with repect to the stellar surface.

\begin{figure*}
\hspace{-2.5cm}
{\mbox{\psboxto(20cm;10cm){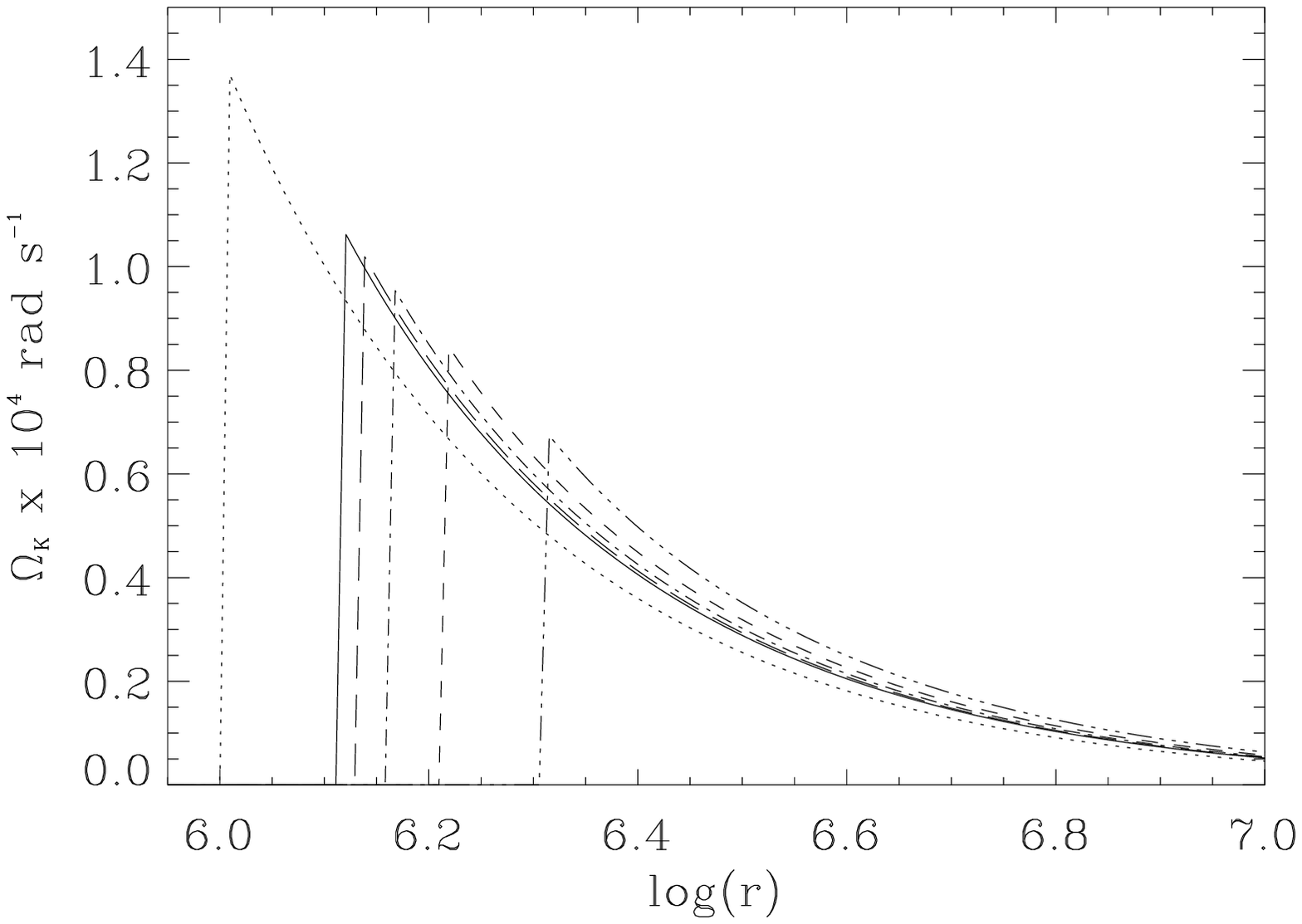}}}
 \caption{ 
 Keplerian angular velocity ($\Omega_K$) profiles  for various EOS.
        The vertical lines indicate the location of the star's surface.
        The horizontal axis corresponds to the logarithm of the radial
        coordinate $r$ taken in centimeters.
}
\end{figure*}

In Fig. 7 we give plots of the angular velocity profiles for 
the radius cross over densities for each of the EOS. 

\section{Discussions}

In this paper we have investigated in  a general relativistic manner, the
effect of rapid rotation on the boundary layer and disk luminosity for
an accreting, old neutron star.  The assumption made is that the magnetic
field of the neutron star is too small to affect the accretion flow.  It
is relevant to ask if a quantitative estimate is possible of how low the 
magnetic
field should be for the validity of our calculations.  The Alfv\'{e}n
radius ($r_A$) is defined by the relationship (see Lamb, Pethick \& Pines
1973)

\begin{eqnarray}
\frac{B^2(r_A)}{8\pi} & = & \rho(r_A)v^2(r_A)
\end{eqnarray}

\noindent where $\rho$ and $v$ are respectively the density and radial 
velocity in the accretion disk.  The Alfv\'{e}n radius determines the
location at which magnetic pressure channels the flow from a disk into
an accretion column structure above the magnetic poles.  Lamb, Pethick \&
Pines (1973) show that

\begin{eqnarray}
r_A & \lteqe & 2.6 \times 10^8 \left[ \frac{\mu_{30}^{4/7}(M/\msune)^{1/7}}
{L^{2/7}_{37} R^{2/7}_6}\right] cm
\end{eqnarray}

\noindent where $\mu_{30} = B_0R^3/10^{30}~G~cm^3$, $L_{37}$ is the total
luminosity in units of $10^{37}~ergs~s^{-1}$, $R_6=R/10^6$ cm and $B_0$ is 
the magnetic field on the surface of the neutron star in gauss.  The condition
that $r_A<R$ implies that (for the reasonable choice: $M=1.4$ \msun and 
$R_6=1$):

\begin{eqnarray}
B_0 & < & 5.5 \times 10^7 L_{37}^{1/2} 
\end{eqnarray}

\noindent and is necessary for the scenario we have considered to be fully
self--consistent. In our notation, $L=(E_D+E_{BL})\dot{M}c^2$, with $\dot{M}$
the mass accretion rate.  

The main conclusions of this paper are that rotational effects of general 
relativity increase the disk luminosity and more importantly from the point 
of observations, decrease the boundary layer luminosity. 
These effects are small in magnitude for small values of $\Omega$ but 
become substantial for rapid rotation rates of the neutron star. 
The boundary layer luminosity becomes inconsequential for rotation rates 
near the centrifugal mass shed limit.  For such cases, the role of radiation
pressure on the accretion flow (see Miller \& Lamb 1996) must be re--examined. 
Also for such cases, accretion induced changes in the surface properties of
the neutron star is an important question to investigate.
The vanishing of the boundary layer luminosity for rapid rotation rates as
found in this study is not apparent in a similar calculation using the
``slow'' rotation approximation based on the Hartle \& Thorne (1968) metric
(see Datta, Thampan \& Wiita 1995).  The total luminosity 
remains fairly constant upto rotation rate of about $0.6 \Omega_{ms}$, 
 but declines rapidly to the value of the disk luminosity for higher 
rotation rates. We have not considered here the angular momentum
evolution of the accreting neutron star, but have calculated the 
accretion luminosities for chosen fixed values of $\Omega$.
 So, the boundary layer luminosity values listed in Table 1
do not include corrections for the energy that may go into the spinning 
up of the neutron star. 
We have not considered the effect of viscosity on
the accretion luminosity.  The viscosity effects require a full
hydrodynamic treatment  and also the radiative transfer phenomenon.  

 An interesting conclusion of the present study (see Fig 4) is
that neutron star configurations with high central densities have 
their innermost stable circular orbit located exterior to the star.  For
such configurations that are rotating at the centrifugal mass shed limit, 
particles in the innermost stable circular orbit  are more bound than particles 
at the surface of the star. This could lead to the formation of an inner disk 
torus. The idea of an inner disk torus has been  invoked as a possible  
explanation of flaring branch phenomena observed in certain Quasi Periodic 
Oscillators (Kuulkers \& van der Klis 1995), with radiation pressure playing 
an key dynamical 
role. Our study seems to suggest that an inner disk torus can be formed 
in the absence of a substantial radiation pressure, purely as a consequence of 
general relativistic rotational space--time in situations where the rotation
rate of the accreting neutron star is close to the centrifugal mass shed limit.

\section*{Acknowledgements} 

 We thank J. Friedman and N. Stergioulas for 
supplying us with their code for rapidly rotating neutron stars, which helped
in checking the accuracy of our code. Thanks are also due to I. Bombaci for
the Baldo--Bombaci--Burgio EOS tables and G. Srinivasan and 
R. Nayak for helpful comments.



%

\bsp

\label{lastpage}
\end{document}

%% file: psbox.tex
%
%
%
%
%
\def\temp{1.34}%
\let\tempp=\relax
\expandafter\ifx\csname psboxversion\endcsname\relax
  \message{PSBOX(\temp) loading}%
\else
    \ifdim\temp cm>\psboxversion cm
      \message{PSBOX(\temp) loading}%
    \else
      \message{PSBOX(\psboxversion) is already loaded: I won't load
        PSBOX(\temp)!}%
      \let\temp=\psboxversion
      \let\tempp= 
    \fi
\fi
\tempp
\let\psboxversion=\temp
\catcode`\@=11
%
%
\def\psfortextures{
\def\PSspeci@l##1##2{%
\special{illustration ##1\space scaled ##2}%
}}%
\def\psfordvitops{
\def\PSspeci@l##1##2{%
\special{dvitops: import ##1\space \the\drawingwd \the\drawinght}%
}}%
\def\psfordvips{
\def\PSspeci@l##1##2{%
\d@my=0.1bp \d@mx=\drawingwd \divide\d@mx by\d@my
\includegraphics{##1\space}}}%
\def\psforoztex{
\def\PSspeci@l##1##2{%
\special{##1 \space
      ##2 1000 div dup scale
      \number-\psllx\space \number-\pslly\space translate
}}}%
\def\psfordvitps{
\def\psdimt@n@sp##1{\d@mx=##1\relax\edef\psn@sp{\number\d@mx}}
\def\PSspeci@l##1##2{%
\special{dvitps: Include0 "psfig.psr"}
\psdimt@n@sp{\drawingwd}
\special{dvitps: Literal "\psn@sp\space"}
\psdimt@n@sp{\drawinght}
\special{dvitps: Literal "\psn@sp\space"}
\psdimt@n@sp{\psllx bp}
\special{dvitps: Literal "\psn@sp\space"}
\psdimt@n@sp{\pslly bp}
\special{dvitps: Literal "\psn@sp\space"}
\psdimt@n@sp{\psurx bp}
\special{dvitps: Literal "\psn@sp\space"}
\psdimt@n@sp{\psury bp}
\special{dvitps: Literal "\psn@sp\space startTexFig\space"}
\special{dvitps: Include1 "##1"}
\special{dvitps: Literal "endTexFig\space"}
}}%
\def\psfordvialw{
\def\PSspeci@l##1##2{
\special{language "PostScript",
position = "bottom left",
literal "  \psllx\space \pslly\space translate
  ##2 1000 div dup scale
  -\psllx\space -\pslly\space translate",
include "##1"}
}}%
\def\psforptips{
\def\PSspeci@l##1##2{{
\d@mx=\psurx bp
\advance \d@mx by -\psllx bp
\divide \d@mx by 1000\multiply\d@mx by \xscale
\incm{\d@mx}
\let\tmpx\dimincm
\d@my=\psury bp
\advance \d@my by -\pslly bp
\divide \d@my by 1000\multiply\d@my by \xscale
\incm{\d@my}
\let\tmpy\dimincm
\d@mx=-\psllx bp
\divide \d@mx by 1000\multiply\d@mx by \xscale
\d@my=-\pslly bp
\divide \d@my by 1000\multiply\d@my by \xscale
\at(\d@mx;\d@my){\special{ps:##1 x=\tmpx, y=\tmpy}}
}}}%
\def\psonlyboxes{
\def\PSspeci@l##1##2{%
\at(0cm;0cm){\boxit{\vbox to\drawinght
  {\vss\hbox to\drawingwd{\at(0cm;0cm){\hbox{({\tt##1})}}\hss}}}}
}}%
\def\psloc@lerr#1{%
\let\savedPSspeci@l=\PSspeci@l%
\def\PSspeci@l##1##2{%
\at(0cm;0cm){\boxit{\vbox to\drawinght
  {\vss\hbox to\drawingwd{\at(0cm;0cm){\hbox{({\tt##1}) #1}}\hss}}}}
\let\PSspeci@l=\savedPSspeci@l
}}%
%
%
\newread\pst@mpin
\newdimen\drawinght\newdimen\drawingwd
\newdimen\psxoffset\newdimen\psyoffset
\newbox\drawingBox
\newcount\xscale \newcount\yscale \newdimen\pscm\pscm=1cm
\newdimen\d@mx \newdimen\d@my
\newdimen\pswdincr \newdimen\pshtincr
\let\ps@nnotation=\relax
{\catcode`\|=0 |catcode`|\=12 |catcode`|
|catcode`#=12 |catcode`*=14
|xdef|backslashother{\}*
|xdef|percentother{
|xdef|tildeother{~}*
|xdef|sharpother{#}*
}%
\def\R@moveMeaningHeader#1:->{}%
\def\uncatcode#1{%
\edef#1{\expandafter\R@moveMeaningHeader\meaning#1}}%
\def\execute#1{#1}
\def\psm@keother#1{\catcode`#112\relax}
\def\executeinspecs#1{%
\execute{\begingroup\let\do\psm@keother\dospecials\catcode`\^^M=9#1\endgroup}}%
\def\@mpty{}%
\def\matchexpin#1#2{
  \fi%
  \edef\tmpb{{#2}}%
  \expandafter\makem@tchtmp\tmpb%
  \edef\tmpa{#1}\edef\tmpb{#2}%
  \expandafter\expandafter\expandafter\m@tchtmp\expandafter\tmpa\tmpb\endm@tch%
  \if\match%
}%
\def\matchin#1#2{%
  \fi%
  \makem@tchtmp{#2}%
  \m@tchtmp#1#2\endm@tch%
  \if\match%
}%
\def\makem@tchtmp#1{\def\m@tchtmp##1#1##2\endm@tch{%
  \def\tmpa{##1}\def\tmpb{##2}\let\m@tchtmp=\relax%
  \ifx\tmpb\@mpty\def\match{YN}%
  \else\def\match{YY}\fi%
}}%
\def\incm#1{{\psxoffset=1cm\d@my=#1
 \d@mx=\d@my
  \divide\d@mx by \psxoffset
  \xdef\dimincm{\number\d@mx.}
  \advance\d@my by -\number\d@mx cm
  \multiply\d@my by 100
 \d@mx=\d@my
  \divide\d@mx by \psxoffset
  \edef\dimincm{\dimincm\number\d@mx}
  \advance\d@my by -\number\d@mx cm
  \multiply\d@my by 100
 \d@mx=\d@my
  \divide\d@mx by \psxoffset
  \xdef\dimincm{\dimincm\number\d@mx}
}}%
%
\newif\ifNotB@undingBox
\newhelp\PShelp{Proceed: you'll have a 5cm square blank box instead of
your graphics (Jean Orloff).}%
\def\s@tsize#1 #2 #3 #4\@ndsize{
  \def\psllx{#1}\def\pslly{#2}%
  \def\psurx{#3}\def\psury{#4}
  \ifx\psurx\@mpty\NotB@undingBoxtrue
  \else
    \drawinght=#4bp\advance\drawinght by-#2bp
    \drawingwd=#3bp\advance\drawingwd by-#1bp
  \fi
  }%
\def\sc@nBBline#1:#2\@ndBBline{\edef\p@rameter{#1}\edef\v@lue{#2}}%
\def\g@bblefirstblank#1#2:{\ifx#1 \else#1\fi#2}%
{\catcode`\%=12
\xdef\B@undingBox{
\def\ReadPSize#1{
 \readfilename#1\relax
 \let\PSfilename=\lastreadfilename
 \openin\pst@mpin=#1\relax
 \ifeof\pst@mpin \errhelp=\PShelp
   \errmessage{I haven't found your postscript file (\PSfilename)}%
   \psloc@lerr{was not found}%
   \s@tsize 0 0 142 142\@ndsize
   \closein\pst@mpin
 \else
   \if\matchexpin{\GlobalInputList}{, \lastreadfilename}%
   \else\xdef\GlobalInputList{\GlobalInputList, \lastreadfilename}%
     \immediate\write\psbj@inaux{\lastreadfilename,}%
   \fi%
   \loop
     \executeinspecs{\catcode`\ =10\global\read\pst@mpin to\n@xtline}%
     \ifeof\pst@mpin
       \errhelp=\PShelp
       \errmessage{(\PSfilename) is not an Encapsulated PostScript File:
           I could not find any \B@undingBox: line.}%
       \edef\v@lue{0 0 142 142:}%
       \psloc@lerr{is not an EPSFile}%
       \NotB@undingBoxfalse
     \else
       \expandafter\sc@nBBline\n@xtline:\@ndBBline
       \ifx\p@rameter\B@undingBox\NotB@undingBoxfalse
         \edef\t@mp{%
           \expandafter\g@bblefirstblank\v@lue\space\space\space}%
         \expandafter\s@tsize\t@mp\@ndsize
       \else\NotB@undingBoxtrue
       \fi
     \fi
   \ifNotB@undingBox\repeat
   \closein\pst@mpin
 \fi
\message{#1}%
}%
%
%
\def\psboxto(#1;#2)#3{\vbox{
   \ReadPSize{#3}%
   \divide\drawingwd by 1000
   \divide\drawinght by 1000
   \d@mx=#1
   \ifdim\d@mx=0pt\xscale=1000
         \else \xscale=\d@mx \divide \xscale by \drawingwd\fi
   \d@my=#2
   \ifdim\d@my=0pt\yscale=1000
         \else \yscale=\d@my \divide \yscale by \drawinght\fi
   \ifnum\yscale=1000
         \else\ifnum\xscale=1000\xscale=\yscale
                    \else\ifnum\yscale<\xscale\xscale=\yscale\fi
              \fi
   \fi
   \divide\pswdincr by 1000 \multiply\pswdincr by \xscale
   \divide\pshtincr by 1000 \multiply\pshtincr by \xscale
   \divide\psxoffset by1000 \multiply\psxoffset by\xscale
   \divide\psyoffset by1000 \multiply\psyoffset by\xscale
   \global\divide\pscm by 1000
   \global\multiply\pscm by\xscale
   \multiply\drawingwd by\xscale \multiply\drawinght by\xscale
   \ifdim\d@mx=0pt\d@mx=\drawingwd\fi
   \ifdim\d@my=0pt\d@my=\drawinght\fi
   \message{scaled \the\xscale}%
 \hbox to\d@mx{\hss\vbox to\d@my{\vss
   \global\setbox\drawingBox=\hbox to 0pt{\kern\psxoffset\vbox to 0pt{
      \kern-\psyoffset
      \PSspeci@l{\PSfilename}{\the\xscale}%
      \vss}\hss\ps@nnotation}%
   \advance\pswdincr by \drawingwd
   \advance\pshtincr by \drawinght
   \global\wd\drawingBox=\the\pswdincr
   \global\ht\drawingBox=\the\pshtincr
   \baselineskip=0pt
   \copy\drawingBox
 \vss}\hss}%
  \global\psxoffset=0pt
  \global\psyoffset=0pt
  \global\pswdincr=0pt
  \global\pshtincr=0pt 
  \global\pscm=1cm 
  \global\drawingwd=\drawingwd
  \global\drawinght=\drawinght
}}%
%
%
\def\psboxscaled#1#2{\vbox{
  \ReadPSize{#2}%
  \xscale=#1
  \message{scaled \the\xscale}%
  \advance\drawingwd by\pswdincr\advance\drawinght by\pshtincr
  \divide\pswdincr by 1000 \multiply\pswdincr by \xscale
  \divide\pshtincr by 1000 \multiply\pshtincr by \xscale
  \divide\psxoffset by1000 \multiply\psxoffset by\xscale
  \divide\psyoffset by1000 \multiply\psyoffset by\xscale
  \divide\drawingwd by1000 \multiply\drawingwd by\xscale
  \divide\drawinght by1000 \multiply\drawinght by\xscale
  \global\divide\pscm by 1000
  \global\multiply\pscm by\xscale
  \global\setbox\drawingBox=\hbox to 0pt{\kern\psxoffset\vbox to 0pt{
     \kern-\psyoffset
     \PSspeci@l{\PSfilename}{\the\xscale}%
     \vss}\hss\ps@nnotation}%
  \advance\pswdincr by \drawingwd
  \advance\pshtincr by \drawinght
  \global\wd\drawingBox=\the\pswdincr
  \global\ht\drawingBox=\the\pshtincr
  \baselineskip=0pt
  \copy\drawingBox
  \global\psxoffset=0pt
  \global\psyoffset=0pt
  \global\pswdincr=0pt
  \global\pshtincr=0pt 
  \global\pscm=1cm
  \global\drawingwd=\drawingwd
  \global\drawinght=\drawinght
}}%
%
\def\psbox#1{\psboxscaled{1000}{#1}}%
\newif\ifn@teof\n@teoftrue
\newif\ifc@ntrolline
\newif\ifmatch
\newread\j@insplitin
\newwrite\j@insplitout
\newwrite\psbj@inaux
\immediate\openout\psbj@inaux=psbjoin.aux
\immediate\write\psbj@inaux{\string\joinfiles}%
\immediate\write\psbj@inaux{\jobname,}%
%
%
\def\toother#1{\ifcat\relax#1\else\expandafter%
  \toother@ux\meaning#1\endtoother@ux\fi}%
\def\toother@ux#1 #2#3\endtoother@ux{\def\tmp{#3}%
  \ifx\tmp\@mpty\def\tmp{#2}\let\next=\relax%
  \else\def\next{\toother@ux#2#3\endtoother@ux}\fi%
\next}%
%
%
\let\readfilenamehook=\relax
\def\re@d{\expandafter\re@daux}
\def\re@daux{\futurelet\nextchar\stopre@dtest}%
\def\re@dnext{\xdef\lastreadfilename{\lastreadfilename\nextchar}%
  \afterassignment\re@d\let\nextchar}%
\def\stopre@d{\egroup\readfilenamehook}%
\def\stopre@dtest{%
  \ifcat\nextchar\relax\let\nextread\stopre@d
  \else
    \ifcat\nextchar\space\def\nextread{%
      \afterassignment\stopre@d\chardef\nextchar=`}%
    \else\let\nextread=\re@dnext
      \toother\nextchar
      \edef\nextchar{\tmp}%
    \fi
  \fi\nextread}%
\def\readfilename{\vbox\bgroup%
  \let\\=\backslashother \let\%=\percentother \let\~=\tildeother
  \let\#=\sharpother \xdef\lastreadfilename{}%
  \re@d}%
%
%
\xdef\GlobalInputList{\jobname}%
\def\psnewinput{%
  \def\readfilenamehook{
    \if\matchexpin{\GlobalInputList}{, \lastreadfilename}%
    \else\xdef\GlobalInputList{\GlobalInputList, \lastreadfilename}%
      \immediate\write\psbj@inaux{\lastreadfilename,}%
    \fi%
    \ps@ldinput\lastreadfilename\relax%
    \let\readfilenamehook=\relax%
  }\readfilename%
}%
\expandafter\ifx\csname @@input\endcsname\relax    
  \immediate\let\ps@ldinput=\input\def\input{\psnewinput}%
\else
  \immediate\let\ps@ldinput=\@@input
  \def\@@input{\psnewinput}%
\fi%
\def\nowarnopenout{%
 \def\warnopenout##1##2{%
   \readfilename##2\relax
   \message{\lastreadfilename}%
   \immediate\openout##1=\lastreadfilename\relax}}%
\def\warnopenout#1#2{%
 \readfilename#2\relax
 \def\t@mp{TrashMe,psbjoin.aux,psbjoint.tex,}\uncatcode\t@mp
 \if\matchexpin{\t@mp}{\lastreadfilename,}%
 \else
   \immediate\openin\pst@mpin=\lastreadfilename\relax
   \ifeof\pst@mpin
     \else
     \errhelp{If the content of this file is so precious to you, abort (ie
press x or e) and rename it before retrying.}%
     \errmessage{I'm just about to replace your file named \lastreadfilename}%
   \fi
   \immediate\closein\pst@mpin
 \fi
 \message{\lastreadfilename}%
 \immediate\openout#1=\lastreadfilename\relax}%
{\catcode`\%=12\catcode`\*=14
\gdef\splitfile#1{*
 \readfilename#1\relax
 \immediate\openin\j@insplitin=\lastreadfilename\relax
 \ifeof\j@insplitin
   \message{! I couldn't find and split \lastreadfilename!}*
 \else
   \immediate\openout\j@insplitout=TrashMe
   \message{< Splitting \lastreadfilename\space into}*
   \loop
     \ifeof\j@insplitin
       \immediate\closein\j@insplitin\n@teoffalse
     \else
       \n@teoftrue
       \executeinspecs{\global\read\j@insplitin to\spl@tinline\expandafter
         \ch@ckbeginnewfile\spl@tinline
       \ifc@ntrolline
       \else
         \toks0=\expandafter{\spl@tinline}*
         \immediate\write\j@insplitout{\the\toks0}*
       \fi
     \fi
   \ifn@teof\repeat
   \immediate\closeout\j@insplitout
 \fi\message{>}*
}*
\gdef\ch@ckbeginnewfile#1
 \def\t@mp{#1}*
 \ifx\@mpty\t@mp
   \def\t@mp{#3}*
   \ifx\@mpty\t@mp
     \global\c@ntrollinefalse
   \else
     \immediate\closeout\j@insplitout
     \warnopenout\j@insplitout{#2}*
     \global\c@ntrollinetrue
   \fi
 \else
   \global\c@ntrollinefalse
 \fi}*
\gdef\joinfiles#1\into#2{*
 \message{< Joining following files into}*
 \warnopenout\j@insplitout{#2}*
 \message{:}*
 {*
 \edef\w@##1{\immediate\write\j@insplitout{##1}}*
\w@{
\w@{
\w@{
\w@{
\w@{
\w@{
\w@{
\w@{
\w@{
\w@{
\w@{\string\input\space psbox.tex}*
\w@{\string\splitfile{\string\jobname}}*
\w@{\string\let\string\autojoin=\string\relax}*
}*
 \expandafter\tre@tfilelist#1, \endtre@t
 \immediate\closeout\j@insplitout
 \message{>}*
}*
\gdef\tre@tfilelist#1, #2\endtre@t{*
 \readfilename#1\relax
 \ifx\@mpty\lastreadfilename
 \else
   \immediate\openin\j@insplitin=\lastreadfilename\relax
   \ifeof\j@insplitin
     \errmessage{I couldn't find file \lastreadfilename}*
   \else
     \message{\lastreadfilename}*
     \immediate\write\j@insplitout{
     \executeinspecs{\global\read\j@insplitin to\oldj@ininline}*
     \loop
       \ifeof\j@insplitin\immediate\closein\j@insplitin\n@teoffalse
       \else\n@teoftrue
         \executeinspecs{\global\read\j@insplitin to\j@ininline}*
         \toks0=\expandafter{\oldj@ininline}*
         \let\oldj@ininline=\j@ininline
         \immediate\write\j@insplitout{\the\toks0}*
       \fi
     \ifn@teof
     \repeat
   \immediate\closein\j@insplitin
   \fi
   \tre@tfilelist#2, \endtre@t
 \fi}*
}%
\def\autojoin{%
 \immediate\write\psbj@inaux{\string\into{psbjoint.tex}}%
 \immediate\closeout\psbj@inaux
 \expandafter\joinfiles\GlobalInputList\into{psbjoint.tex}%
}%
%
%
%
\def\centinsert#1{\midinsert\line{\hss#1\hss}\endinsert}%
\def\psannotate#1#2{\vbox{%
  \def\ps@nnotation{#2\global\let\ps@nnotation=\relax}#1}}%
\def\pscaption#1#2{\vbox{%
   \setbox\drawingBox=#1
   \copy\drawingBox
   \vskip\baselineskip
   \vbox{\hsize=\wd\drawingBox\setbox0=\hbox{#2}%
     \ifdim\wd0>\hsize
       \noindent\unhbox0\tolerance=5000
    \else\centerline{\box0}%
    \fi
}}}%
%
\def\at(#1;#2)#3{\setbox0=\hbox{#3}\ht0=0pt\dp0=0pt
  \rlap{\kern#1\vbox to0pt{\kern-#2\box0\vss}}}%
%
\newdimen\gridht \newdimen\gridwd
\def\gridfill(#1;#2){%
  \setbox0=\hbox to 1\pscm
  {\vrule height1\pscm width.4pt\leaders\hrule\hfill}%
  \gridht=#1
  \divide\gridht by \ht0
  \multiply\gridht by \ht0
  \gridwd=#2
  \divide\gridwd by \wd0
  \multiply\gridwd by \wd0
  \advance \gridwd by \wd0
  \vbox to \gridht{\leaders\hbox to\gridwd{\leaders\box0\hfill}\vfill}}%
%
\def\fillinggrid{\at(0cm;0cm){\vbox{%
  \gridfill(\drawinght;\drawingwd)}}}%
%
%
\def\textleftof#1:{%
  \setbox1=#1
  \setbox0=\vbox\bgroup
    \advance\hsize by -\wd1 \advance\hsize by -2em}%
\def\textrightof#1:{%
  \setbox0=#1
  \setbox1=\vbox\bgroup
    \advance\hsize by -\wd0 \advance\hsize by -2em}%
\def\endtext{%
  \egroup
  \hbox to \hsize{\valign{\vfil##\vfil\cr%
\box0\cr%
\noalign{\hss}\box1\cr}}}%
%
\def\frameit#1#2#3{\hbox{\vrule width#1\vbox{%
  \hrule height#1\vskip#2\hbox{\hskip#2\vbox{#3}\hskip#2}%
        \vskip#2\hrule height#1}\vrule width#1}}%
\def\boxit#1{\frameit{0.4pt}{0pt}{#1}}%
\catcode`\@=12 
%
 \psfordvips   